\documentclass{metalcone}
\usepackage[super,comma,sort&compress]{natbib}
\usepackage[T1]{fontenc}
\usepackage[flushleft]{threeparttable}
\usepackage{hyperref}
\hypersetup{colorlinks,citecolor=blue}
\usepackage{mathptmx}
\usepackage{multirow}
\usepackage{amsmath}

\begin{document}
\setlength{\parskip}{1em}
\pagestyle{plain}

\justifying
\title{Experimental observation of ballistic to diffusive transition in AlN thin films}

\maketitle


\author{Md Shafkat Bin Hoque, Michael E. Liao, Saman Zare, Zeyu Liu, Yee Rui Koh, Kenny Huynh, Jingjing Shi, Samuel Graham, Tengfei Luo, Habib Ahmad, W. Alan Doolittle, Mark S. Goorsky, and Patrick E. Hopkins}

\begin{affiliations}
\normalsize Md Shafkat Bin Hoque, Yee Rui Koh, Saman Zare\\
Department of Mechanical and Aerospace Engineering, University of Virginia, Charlottesville, Virginia 22904, USA\hfill

\normalsize Michael E. Liao, Kenny Huynh, Mark S. Goorsky\\
Department of Materials Science and Engineering, University of California, Los Angeles, California 90095, USA\hfill

\normalsize Zeyu Liu\\
Department of Applied Physics, Hunan University, Changsha 410300, China\hfill

\normalsize Jingjing Shi\\
George W. Woodruff School of Mechanical Engineering, Georgia Institute of Technology, Atlanta, Georgia 30332, USA\hfill

\normalsize Samuel Graham\\
George W. Woodruff School of Mechanical Engineering, Georgia Institute of Technology, Atlanta, Georgia 30332, USA\hfill
School of Materials Science and Engineering, Georgia Institute of Technology, Atlanta, Georgia 30332, USA\hfill

\normalsize Tengfei Luo\\
Department of Aerospace and Mechanical
Engineering, University of Notre Dame, Notre Dame, Indiana
46556, USA\hfill

\normalsize Habib Ahmad, W. Alan Doolittle\\
School of Electrical and Computer Engineering, Georgia Institute of Technology, Atlanta, Georgia 30332, USA\hfill

\normalsize Patrick E. Hopkins\\
Department of Mechanical and Aerospace Engineering, University of Virginia, Charlottesville, Virginia 22904, USA\\
Department of Materials Science and Engineering, University of Virginia, Charlottesville, Virginia 22904, USA\\
Department of Physics, University of Virginia, Charlottesville, Virginia 22904, USA\\
Email: phopkins@virginia.edu

\end{affiliations}


\dedication{}

\dedication{}


\newpage
\begin{abstract}

Bulk AlN possesses high thermal conductivity due to long phonon mean-free-paths, high group velocity, and long lifetimes. However, the thermal transport scenario becomes very different in a thin AlN film due to phonon-defect and phonon-boundary scattering. Herein, we report experimental observation of ballistic to diffusive transition in a series of AlN thin films (1.6 - 2440 nm) grown on sapphire substrates. The ballistic transport is characterized by constant thermal resistance as a function of film thickness due to phonon scattering by defects and boundaries. In this transport regime, phonons possess very small group velocities and lifetimes. The lifetime of the optical phonons increases by more than an order of magnitude in the diffusive regime, however, remains nearly constant afterwards. Our study is important for understanding the details of nano and microscale thermal transport in a highly conductive material.     

\end{abstract}



\section{Introduction}
In the last half-century, the dimensions of electronic, photonic, and optoelectronic devices have decreased to nano and micrometer length scales.\cite{prasher2005nano,pop2006heat,shoemaker2023implications} One of the major bottlenecks to the efficient operation and long lifetime of functional devices is non-uniform heat dissipation, resulting in hot-spot generation.\cite{oh2019thermal,he2021thermal} One solution is passive cooling by using highly conductive crystalline materials as interfacial layers and substrate materials.\cite{mahajan2006cooling,pop2010energy,moore2014emerging} Ultrawide bandgap semiconductor aluminum nitride (AlN) is quite promising in this regard because of its high thermal conductivity and requisite electronic properties, making it suitable for high electron mobility transistors, ultraviolet light emitting diodes, and high power radio frequency devices.\cite{miyashiro1990high,lindsay2013ab,koehler2017ultrawide} While several studies have reported the thermal properties of micrometer thick AlN films,\cite{cheng2020experimental,koh2020bulk,hoque2021high} thermal characterizations of nanoscale AlN thin films are almost non-existent in literature. Thermal transport in nanoscale AlN films is therefore not well understood.

Thin film materials can exhibit two regimes of thermal transport: ballistic and diffusive.\cite{wang2006carbon,maldovan2012transition,giri2021thickness} In the ballistic regime, thermal resistance remains nearly constant as a function of film thickness. This regime is observed when the mean-free-path of energy carriers is much larger than the thin film dimensions. Several molecular dynamics (MD) studies have predicted ballistic transport in interfacial confined films.\cite{landry2010effect,giri2016effect} On the other hand, in the diffusive regime, cross-plane thermal resistance increases with thickness. As the phonon mean-free-path (MFP) of AlN is nearly $\sim$3 $\mu$m at room temperature,\cite{hoque2021high} possibility exists for observing ballistic transport in nano and microscale AlN films. Such transport can reduce thermal resistance and enable better heat dissipation. Therefore, understanding ballistic and diffusive transport regimes in AlN films is important for choosing the optimum film thickness for functional device geometry.

Motivated by this, we study a series of AlN films grown via the metal modulated epitaxy (MME) technique. The AlN films are sandwiched between an aluminum (Al) film and a sapphire substrate and have thicknesses ranging from 1.6 to 2440 nm. Transmission electron microscopy (TEM) reveals that the microstructure of the AlN films changes drastically as the thickness increases. Time-domain thermoreflectance (TDTR) is used to measure the thermal resistance across the Al/AlN/sapphire interface. We observe ballistic transport when the film thickness is less than 30 nm; beyond this thickness, thermal transport becomes diffusive. We attribute the ballistic transport to long MFP heat carrying phonons being scattered by film boundaries and defects. Boltzmann transport equation (BTE) calculations show that the majority of the phonon modes possess a small group velocity in the ballistic regime. We further use the IR-VASE technique to measure the lifetime of the optical phonons. Our measurements reveal that the optical phonons have very small lifetimes in the ballistic regime, however, they increase by an order of magnitude in the diffusive regime. Interestingly, the lifetime remains nearly constant as the film thickness increases from 119 to 2440 nm. We therefore attribute the thermal conductivity accumulation in diffusive regime to acoustic phonons. Our study provides insight into thermal transport of AlN films and enables choosing optimum thicknesses for functional devices.

\section{Growth Details of the AlN films}

The sapphire substrates are first subjected to pre-treatment for cleaning and achieving atomically flat sample surface. Details of the surface pre-treatment are provided in the Supporting Information. The substrates are then loaded into an introductory chamber at a vacuum level of $\sim$10$^{-9}$ Torr. They are thermally outgassed for \textit{in situ} thermal cleaning first in the introductory vacuum chamber at 200 $^\circ$C for 20 minutes and subsequently outgassed in the growth chamber at 850 $^\circ$C for 10 minutes. 

\textit{In situ} nitridation of the sapphire substrates is later performed inside the molecular beam epitaxy (MBE) chamber at 6.3 sccm nitrogen flow rate, 350 W plasma power, and at 200 $^\circ$C substrate temperature. The full nitridation time is 25 min. A Veeco UNI-bulb radio frequency nitrogen plasma source is employed for this purpose.

The AlN films are grown via metal modulated epitaxy (MME) at a RF plasma power of 350 W and a flow rate of 2.5 sccm. The RF plasma power and flow rates are kept uniform for all growths. The MBE growth chamber base pressure is $\sim$5 × 10$^{-11}$ Torr and the beam equivalent pressure of the nitrogen plasma is $\sim$2 × 10$^{-5}$ Torr. The AlN films are grown at a substrate temperature of 800 $^\circ$C. The $\sim$80 nm aluminum films are deposited at 200 $^\circ$C substrate temperature for thermal measurements.

\section{Characterizations of the AlN films}

\begin{figure}[hbt!]
\centering
\includegraphics[scale = 0.4]{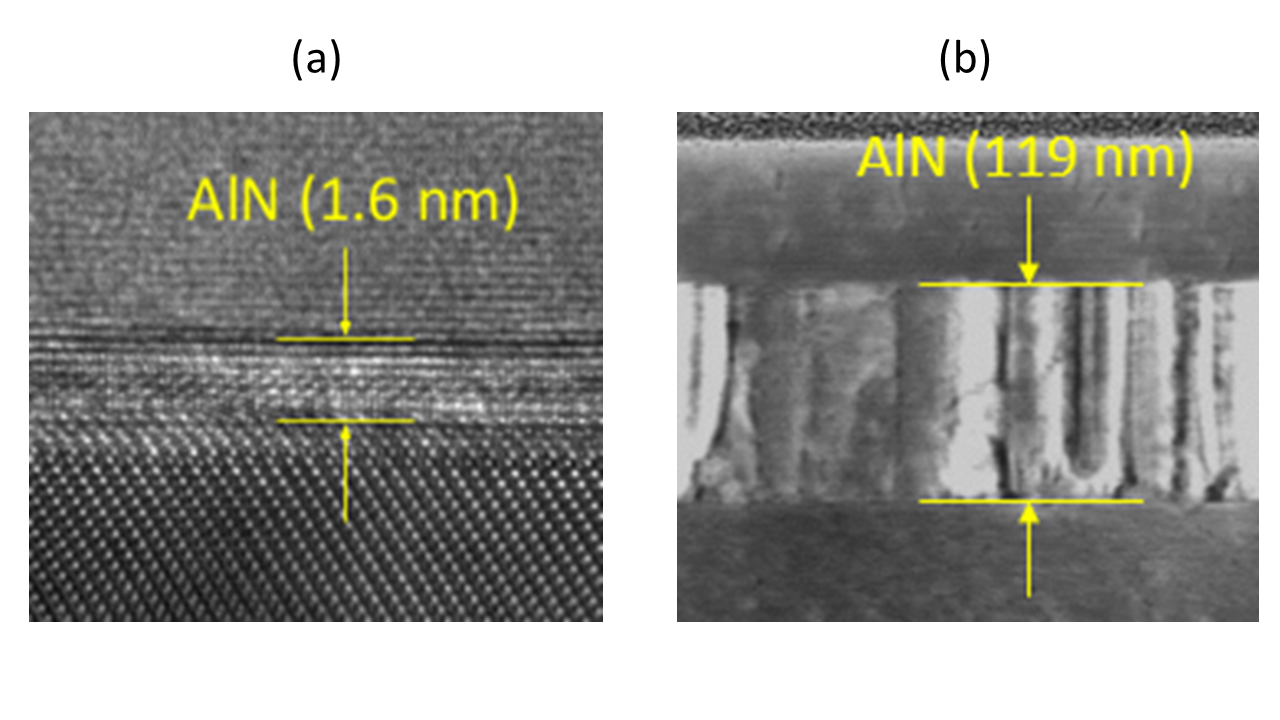}\\
\includegraphics[scale = 0.4]{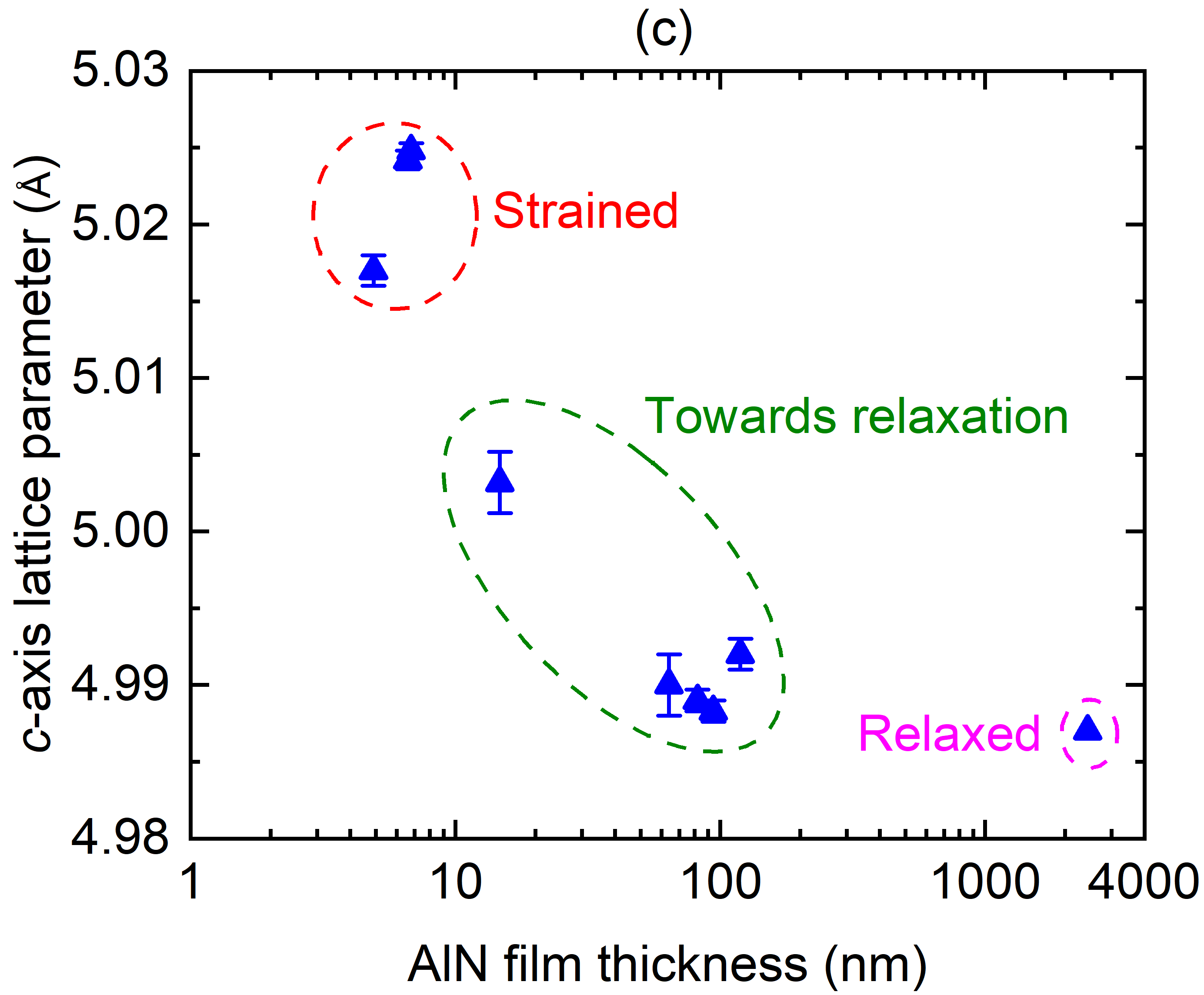}\\
\caption{TEM characterizations of (a) 1.6 and (b) 119 nm AlN thins films. (c) \textit{c}-axis lattie parameter as a function of film thickness.}
\label{fig:6_2}
\end{figure}

We use transmission electron microscopy (TEM) to characterize the microstructure of the thin films. Figures 1(a) and 1(b) shows the TEM characterizations of 1.6 and 119 nm AlN films, respectively. As the film thickness increases from 1.6 to 119 nm, columnar grains are introduced in the films.   

The strain in the AlN thin films also changes greatly as measured by the  \textit{c}-axis lattice parameter in Figure 1(c). Nitridation of the (0001) sapphire surface initially produces a highly strained uniform 1.6 nm thin AlN layer. The sapphire substrate imposes in-plane biaxial compressive stress on the AlN, which induces out-of-plane tensile strain. The maximum tensile strain is observed for thin layers up to 10 nm of AlN. As the AlN thickness increases, strain-relieving misfit dislocations form and consequently relax the AlN film. AlN relaxation monotonically increases with increasing thickness up to 100 nm of AlN as seen by the decrease in the out-of-plane \textit{c}-axis lattice parameter. The AlN film is fully relaxed for thicknesses greater than 1 $\mu$m. The strain change is also an indicator of microstructural changes within the AlN films.

\section{Results and discussions}

Optical pump-probe technique time-domain thermoreflectance (TDTR) is used for thermal characterizations of the samples. We use co-axially focused 1/e$^2$ pump and probe diameters $\sim$19 and 10 $\mu$m, respectively. Details of our two-tint TDTR setup can be found elsewhere.\cite{jiang2018tutorial,olson2020anisotropic,olson2021evolution,hoque2024disorder} 

The TDTR-measured thermal resistance (\textit{R}) across the Al/AlN/sapphire interface can be expressed as the following:

\begin{equation}
\textit{R} = \bigg(\frac{1}{G}\bigg)_{Al/AlN} + \bigg(\frac{L}{\kappa}\bigg)_{AlN} + \bigg(\frac{1}{G}\bigg)_{AlN/Sapphire}
\end{equation}

Where \textit{G}, $\kappa$, and \textit{L} represent thermal boundary conductance, thermal conductivity, and thickness of the AlN films, respectively. We set Al/AlN and AlN/sapphire thermal boundary conductances to infinity\cite{cancellieri2020interface,lorenzin2022tensile} and fit for the AlN thin film thermal conductivity. 

Figure 2(a) shows the thermal resistance as a function of AlN film thickness. As exhibited here, when the film thickness is less than 30 nm, thermal resistance remains nearly constant, i.e., transport is ballistic. This thermal resistance value is slightly higher compared to a Al/sapphire interfacial resistance.\cite{koh2020thermal} Beyond the 30 nm thickness, thermal resistance increases with thickness, indicating a transition to diffusive regime. 

To gain more insight into this, we plot the AlN thermal conductivity accumulation as a function of phonon MFP. The MFP is calculated by solving linearized phonon Boltzmann transport equation (BTE) with the help of first-principles force constants.\cite{hoque2021high} When the MFP is 30 nm or less, the thermal conductivity accumulation is almost zero. This indicates that phonons with MFPs less than 30 nm have negligible contribution to thermal conductivity. The excellent agreement between experimental data and calculations show that when the AlN film thickness is less than 30 nm, heat carrying long MFP phonons are scattered by the AlN/sapphire interface and structural defects. The evolution of microstructural defects within the AlN films as a function of thickness is visible in Figure 1(c). In absence of defects, the thermal resistance would slightly decrease in the ballistic regime as shown by the MD simulations in Supporting Figure S1. As the film thickness becomes greater than 30 nm, long MFP phonons start to contribute to thermal transport along the cross-plane direction, leading to diffusive regime. 

\begin{figure}[hbt!]
\centering
\includegraphics[width=\textwidth]{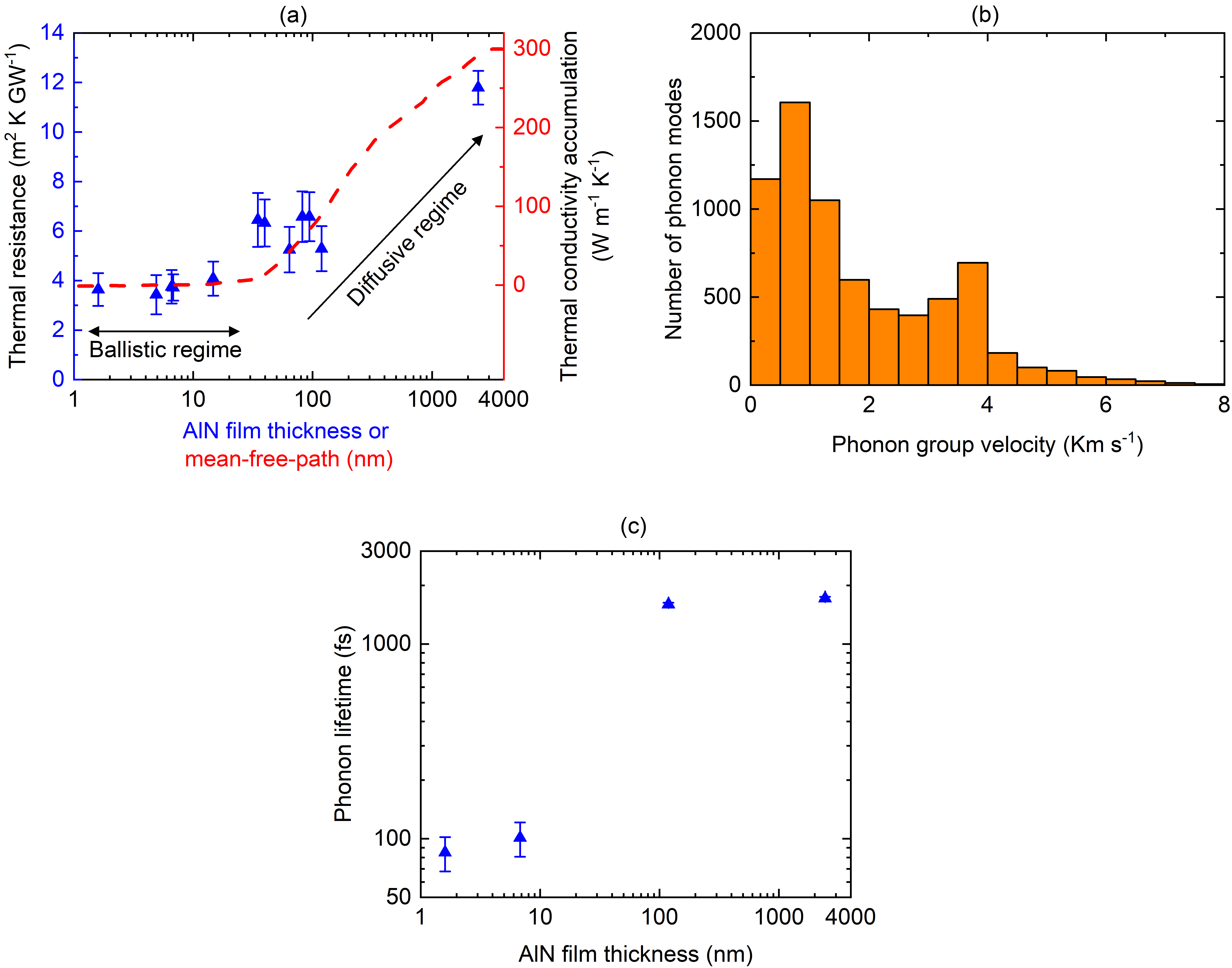}\\
\caption{(a) Thermal resistance as a function of AlN film thickness, and thermal conductivity accumulation as a function of phonon mean-free-path. (b) Number of phonon modes vs group velocity for phonons with mean-free-paths less than 30 nm. (c) Phonon lifetimes measured by the IR-VASE technique as a function of AlN film thickness. }
\label{fig:6_2}
\end{figure}

The phonon MFP is related to two other parameters: group velocity and lifetime. For a qualitative analysis, we calculate the group velocity distribution of phonons with MFP less than 30 nm and plot it in Figure 2(b). The BTE calculations here considers a thickness dependent scattering rates and MFP. As observed here, the major phonon modes have a very small group velocity, with the averaged group velocity being 1818 m s$^{-1}$. For comparison, the group velocity of bulk AlN is $\sim$7000 m s$^{-1}$.\cite{gerlich1986elastic,sahoo2012effect,aryana2021tuning} This shows the phonon group velocity is very small in the ballistic regime.

We use the IR-VASE technique to measure the lifetime of the optical phonons in the AlN films. Details of the IR-VASE measurements can be found in previous publication.\cite{hoque2024disorder} The lifetimes remain nearly constant in both ballistic and diffusive regime, and exhibits an order of magnitude change across ballistic to diffusive transition. In the ballistic regime, optical phonon lifetimes are quite small. This stems from the extensive phonon-defect and phonon-boundary scattering in this regime. The near constant lifetime in the diffusive regime indicates that film thickness or boundary scattering is not impacting the optical phonons in this regime. We therefore attribute the thermal conductivity accumulation of Figure 2(a) to acoustic phonons. 

\section{Conclusion}

In summary, we report on experimental observations of ballistic to diffusive transition in AlN thin films. The crystal structure and strain within the AlN films change significantly as the thickness increases from 1.6 to 2440 nm. The ballistic to diffusive transition occurs when the film thickness is $\sim$30 nm. In the ballistic regime, phonons posses small group velocities and lifetimes. The optical phonon lifetimes increase significantly across the ballistic to diffusive transition point. Our study provides a physical picture of thermal transport in nano and microscale AlN thin films.\\

\textbf{Acknowledgement}

The authors acknowledge financial support from the Office of Naval Research Grant Number N00014-23-1-2630.

\newpage
\textbf{Supporting Information}\\

\textbf{S1. Surface pre-treatment of sapphire substrates}

High temperature annealing is an effective method to achieve atomically flat surface of sapphire substrates. High quality films can be grown on top of annealed sapphire substrates. The \textit{c}-plane sapphire substrates are first \textit{ex situ} annealed at 1175 $^\circ$C via a multi-step annealing procedure. For annealing, the sapphire wafers are placed in a quartz boat and pre-heated at 1075 $^\circ$C for 5 minutes under nitrogen environment. The wafers are then moved to the center furnace zone and heated at 1075 $^\circ$C under nitrogen environment for 15 minutes. The wafers are subsequently heated under ultra zero grade air environment at 1075 $^\circ$C for 1 hour. Temperature is ramped up to 1175 $^\circ$C and the sapphire wafers are annealed at this temperature under ultra zero grade air for 5 hours. Subsequently, the wafers are allowed to naturally cool down under nitrogen.

Atomic force microscopy (AFM) of the un-annealed sapphire wafer shows disorderly corrugations on the substrate surface with a surface roughness of 0.4 nm rms, while AFM of the annealed sapphire wafers have a surface roughness of approximately 0.05 nm. This shows an order of magnitude improvement in the surface smoothness as compared to the un-annealed wafer. In contrast to the random surface of the un-annealed sapphire, a distinct terrace-and-step structure is also observed on the annealed sapphire wafers with terrace widths of 550 nm.

The annealed sapphire wafers are first piranha (3:1 volume ratio of H$_2$SO$_4$:H$_2$O$_2$) cleaned for 1 minute at 150 $^\circ$C followed by 5:1 volume ratio of deionized water to hydrofluoric acid (DI H$_2$O:HF) for 30 seconds. The cleaned wafers are then backside metalized with 2 $\mu$m tantalum for uniform heating during growth. The backside metalized wafers are later diced into 1 cm × 1 cm templates for growth inside the MBE chamber. The metalized and diced templates are subsequently solvent cleaned (acetone clean at 45 $^\circ$C for 20 minutes, 3 minutes methanol clean, DI water clean, and dried with nitrogen) followed by a more rigorous 10 minutes piranha (3:1 volume ratio of H$_2$SO$_4$:H$_2$O$_2$) clean at 150 $^\circ$C to remove organic solvents. The templates are then \textit{ex situ} chemically cleaned in a 10:1 volume ratio of DI H$_2$O:HF for 30 seconds to partially remove the surface oxides followed by DI water rinse and nitrogen dry procedure. 

\newpage
\textbf{S2. Molecular dynamics (MD) simulations of thermal resistance}

Figure S1 shows the MD simulations of thermal resistance across a Al/AlN/sapphire interface vs. AlN film thickness. MD simulations only consider phonon-boundary scattering and exclude phonon-defect scattering.

\begin{figure}[hbt!]
\centering
\renewcommand{\thefigure}{S1}
\includegraphics[scale = 0.5]{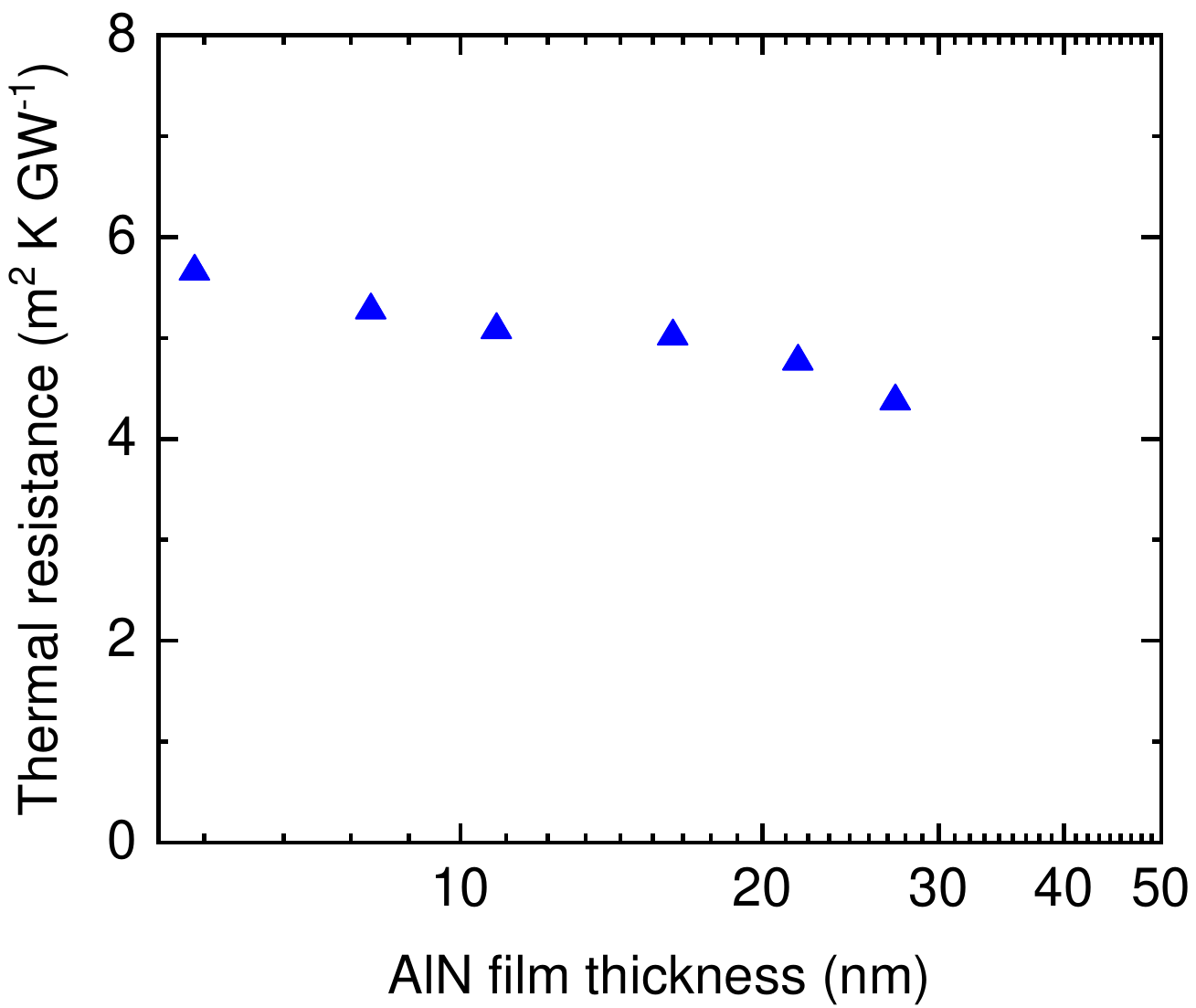}\\
\caption{MD simulations of thermal resistance as a function of AlN film thickness.}
\label{fig:6_2}
\end{figure}

\medskip

%
\newpage
\bibliographystyle{achemso}
\bibliography{References}




\end{document}